\documentclass[%
 reprint,
showpacs,
 amsmath,amssymb,
 aps,
]{revtex4-1}

\usepackage{fancyhdr}
\usepackage{graphicx}
\usepackage{dcolumn}
\usepackage{bm}

\usepackage[none]{hyphenat}

\newcommand{\beginsupplement}{%
        \setcounter{table}{0}
        \renewcommand{\thetable}{S\arabic{table}}%
        \setcounter{figure}{0}
        \renewcommand{\thefigure}{S\arabic{figure}}%
     }

\begin{document}

\title{Hyperbolic Quantum Processor.}%

\author{Evgenii E. Narimanov$^1$ and Eugene A. Demler$^2$}
\affiliation{$^1$School of Electrical and Computer Engineering and Birck Nanotechnology Center, Purdue University, West Lafayette, Indiana 47907, USA}
\affiliation{$^2$Institute for Theoretical Physics, ETH Zurich, Wolfgang-Pauli-Str. 27, 8093 Zurich, Switzerland}
\date{\today}

\begin{abstract}

Achieving strong coherent interaction between qubits separated by large distances holds the key to many important developments in quantum technology, including new designs of quantum computers,  \cite{Kimble2008,CiracZoller1995,PorrasCirac2004} new platforms for quantum simulations \cite{HungKimble2016} and implementation of large scale quantum optical networks \cite{Arrazola2021}. However,  the inherent mismatch between the spatial dimensions of a quantum emitter and the photon wavelength \cite{Cai} fundamentally limits the transmission of quantum entanglement over long distances. Here we demonstrate, that long-range 
qubit entanglement can be readily achieved when qubit interactions are mediated by optical polariton
waves in a hyperbolic material, due to the phenomenon of the Hyperbolic Super-Resonance. We show that in this regime the resulting quantum gate fidelity that exceeds 99\%, can be achieved with the use of qubits based on well known deep donors in silicon \cite{JanzenGrimmeiss1984,StegerAger2009,AbrahamThewalt2018,KaraiskajCardona2003,SwartzThomas19800}  when their interactions are
mediated by polariton fields in the substrate formed by a hyperbolic material (such as e.g. hexagonal boron nitride \cite{hBN1,DaiBasov2015,hBN2,hBN3,hBN4,YoxallHillenbrand2015,LowKoppens2017,BasonFoglerDeAbajo2016,NiBasov2021,SternbachBasov2023}). At the physical level the proposed system is essentially a silicon-based optoelectronic chip, and it's readily accessible to the existing methods of semiconductor nanofabrication, leading to the integration densities of well over $10^8$ 
qubits/${\rm cm}^{2}$, and therefore opening the way to  scalable and fault-tolerant error correction in quantum computation.
 Furthermore, we demonstrate that, due to the optical time scales that define the duration of the gate operation in the proposed system, and sub-nanosecond time of the decoherence  in deep donors in silicon at the 
liquid nitrogen temperatures,\cite{QC6}  the proposed Hyperbolic Quantum  Processor does not require dilution refrigeration and therefore offers a pathway to bring quantum computation to the realm of conventional engineering.
\end{abstract}

\maketitle

\section{Introduction}

A hyperbolic electromagnetic medium is a strongly anisotropic natural material or a composite structure with the real parts of its dielectric permittivity tensor components showing opposite signs in two orthogonal 
directions.\cite{PN} In artificial (meta)materials, this can be readily achieved using 
extended conducting inclusions (from planar layers \cite{nmat,Vinod} to nanowires \cite{Noginov,Zhang} to elongated nanoparticles \cite{PN}) in a dielectric host, as long as the corresponding unit cell size is well below the free space wavelength.\cite{Cai}  In natural materials, the hyperbolic 
electromagnetic response  can arise from a strong anisotropy in the electronic \cite{Bi} or optical phonon \cite{Sapphire1,Sapphire2,hBN1,hBN2} spectra. 

The transition to a hyperbolic environment dramatically  changes the nature of the propagating waves in the material, whose wavenumbers are no longer limited by the frequency.\cite{hyperlens1} As a result, hyperbolic materials are no longer subject to the diffraction limit on the optical resolution \cite{hyperlens1,hyperlens2} leading to practical realizations of super-resolution imaging \cite{hyperlensE1,hyperlensE2} and sub-diffraction focusing,\cite{DaiBasov2015} and support a broadband singularity of the photonics density of states \cite{PDOSprl} that dramatically changes the quantum electrodynamics \cite{PDOSapl,ScienceTT} in these media. Most importantly, electromagnetic field in a hyperbolic medium is therefore free  from the ``light-matter interaction bottleneck''  between the 
 the spatial dimensions of the quantum emitter and the optical photon wavelength, and can probe the regime of ultra-strong coupling. 

Recent advances in material fabrication brought in the development of strongly anisotropic polar crystals with hyperbolic response in the Reststrahlen bands. \cite{hBN1,DaiBasov2015,hBN2,hBN3,hBN4,YoxallHillenbrand2015,LowKoppens2017,BasonFoglerDeAbajo2016,NiBasov2021,SternbachBasov2023,WangLow2024,SternbachBasov2021,JiaAsgari2022}
Furthermore, the already record-low loss  in the hyperbolic bands in this materials can be dramatically reduced even further by
isotope enrichment \cite{hBNisotope} that removes photon-on-impurity scattering, and by lowering the system temperature that ``freezes out"  the phonon-phonon scattering.\cite{Ziman}  The polar crystals with hyperbolic response therefore offer an a powerful  platform with extreme light-matter coupling of quantum emitters, with unique capabilities for photonic state control and quantum information processing. 

\begin{figure*}[htbp] 
 \centering
 \includegraphics[width=6.5in]{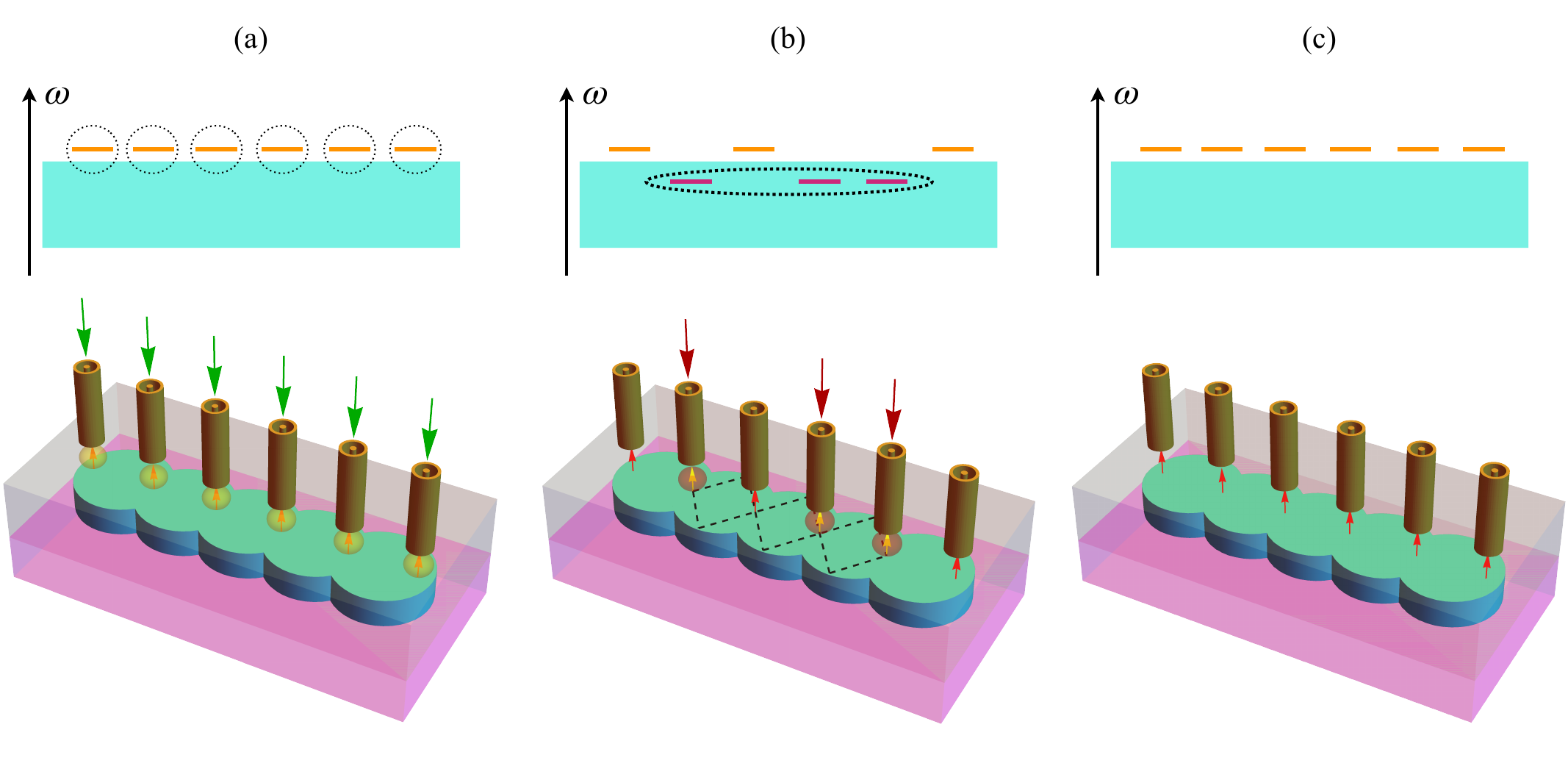} 
   \caption{ Hyperbolic Quantum Processor operation. Panel (a): the quantum transition frequencies $\omega_0$ for all cubits are beyond the hyperbolic frequency band of the hBN. Control fields at $\omega_0$ in the nano-waveguides coupled to selected qubits (green ``clouds''), force the desired single qubit operations. 
  Panel (b): off-resonance control fields (red clouds) applied to selected qubits, down-shift the qubit transition frequencies to the hyperbolic band, which leads to strong dipole-dipole interactions and the resulting entanglement. Panel (c): the off-resonant control fields are released, and the entangled system is ready for the next  quantum gate operation.
   }
   \label{fig:HQPaction}
 \end{figure*}

Furthermore,  due to the phenomenon of hyperbolic Super-Resonance (hSR) that we uncover in this work, even for  atomic-scale quantum emitters such as chalcogen double donors in silicon with long 
quantum coherence (e.g. sulfur,\cite{JanzenGrimmeiss1984,StegerAger2009}, magnesium \cite{Ludwig1965,KlevermanGrimmeiss1986,KaraiskajCardona2003} and selenium\cite{SwartzThomas1980,StegerAger2009}),  the corresponding dipole-dipole interaction energies at spatial separation beyond $100$ nm, when mediated by the high-quality hyperbolic materials such as isotopically enriched hexagonal Boron Nitride,\cite{hBNisotope} show the values in excess of   $30$ meV, which are above $15$\% of the corresponding photon energy( and exceed  the relevant decoherence rates by more than three orders of magnitude.\cite{VinhPNAS2008,GreenlandAeppli2010} Just as importantly, both the photon energy and the coupling strength are well above the thermal energy even at 
room temperature of $22$ meV, so that at the liquid nitrogen conditions any thermal excitations in this system are completely frozen out -- which is essential for 
coherent coupling between the quantum emitters.

With the extreme spatial range of this hyperbolic-mediated interaction approaching the optical wavelength, this allows for a network of physically separate nanophotonic waveguides with minimal cross-talk,
connecting to individual quantum emitters -- which defines the proposed Hyperbolic Quantum Processor (HQP) capable of operation at liquid nitrogen temperatures.

Moreover, since  at  the physical level the HQP is essentially a silicon-based optoelectronic chip, and it's readily accessible to the existing methods of semiconductor nanofabrication, offering the potential for the the integration densities of well over $10^8$ -- it can bring the dream of the scalable fault-tolerant quantum computing to the realm of   practical optoelectronic engineering with today's fabrication technology.

\section{Quantum Computation with Hyperbolic Nanostructures}

On the conceptual level, the proposed Hyperbolic Quantum Processor will use the qubits based on donor atoms  with the optical transition frequency just beyond the hyperbolic band of the hyperbolic waveguide. With the use
of low-loss hyperbolic materials based on high quality polar crystals (such as e.g. sapphire, quartz or hBN),
and appropriate choice of the sickness of the (silicon) spacer between the qubit impurity atom and the interface with the hyperbolic medium,  the additional decoherence  induced by optical absorption is negligible in comparison to  natural transition linewidth, preserving the essential coherence in these donor atoms.

With the free-space optical wavelength in the mind-IR range and the macroscopic qubit separation, a  nanophotonic waveguide network can be readily fabricated (using e.g. the nanoplasmonic waveguides \cite{nano-waveguide}) with desired performance and complexity, that can individually address separate qubits, as shown in Fig. \ref{fig:HQPaction}. Here, single qubit operations can be 
realized in the ``decoupled'' regime when the quantum transition frequency is outside the hyperbolic band -- Fig. \ref{fig:HQPaction}(a),(c).
A desired multi-gate operation can then be achieved by inducing optical Stark shifts by off-resonance  illumination of the chosen qubits./ With such ``control'' fields properly offset from the transition frequency between the ground and excited states of the donor atom, the frequency-shifted atomic transition  will enter the hyperbolic band and tune to the Hyper-Resonance frequency, which results in the strong optical coupling of the chosen qubits and leads to the desired entanglement.  Once the gate operation is complete, the off-resonance control fields are removed, and the cubit transition frequencies are pulled back outside the hyperbolic band, away from the additional decoherence due to hSR-enhanced material absorption in the hyperbolic substrate  -- Fig. \ref{fig:HQPaction}(d).

Note that the general approach of the electromagnetic control of the quantum transition frequencies of individual qubits 
in order to the tune on and off the inter-qubit interactions, is successfully used in superconducting systems.\cite{YanOliver2018,MundadaHouck2019,FoxenMartinis2020,CollodoEichler2020,McKayGambetta2016,SungOliver2021,NegirneacDiCarlo2021}

\begin{figure}[htbp] 
 \centering
 \includegraphics[width=3.5in]{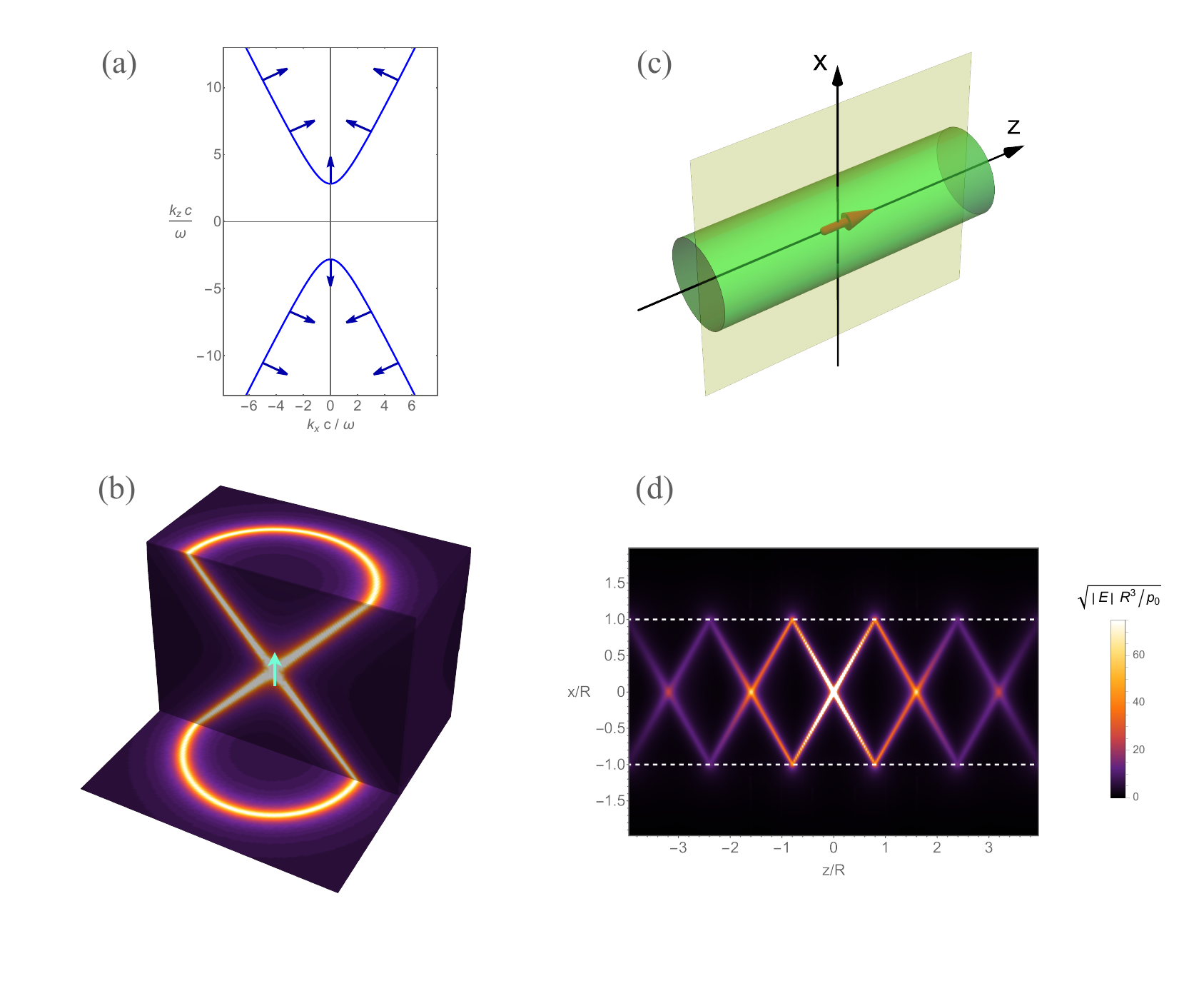} 
   \caption{ Light from a point dipole emission in a hyperbolic medium. Panel (a): the iso-frequency surface for a hyperbolic material with ${\rm Re}[\epsilon_\perp] > 0, \ {\rm Re}[\epsilon_\parallel] < 0$; the arrows show the directions of the
   the group velocity. Panel (b): a cross-cut view of the emission intensity (in false color) of a point dipole (green arrow) in an infinite hyperbolic medium. 
 Panel (c): the schematics of a point dipole emitter at the axis on an infinite cylinder waveguide fabricated from a hyperbolic medium.
   Panels (d): the resulting emission intensity in the waveguide, in false color. $p_0$ is the dipole moment, $R$ is the radius of the cylinder. The material parameters for (b) and (d) correspond to the higher-frequency hyperbolic band in hexagonal boron nitride \cite{hBN1} -- see Fig. \ref{fig:hBN_emitters}.
   }
   \label{fig:HM_dipole}
 \end{figure}

\section{Light in Hyperbolic Media}

In a uniaxial medium, TM-polarized propagating waves are characterized by the dispersion
\begin{eqnarray}
{\epsilon_\parallel} {k_\parallel^2} + {\epsilon_\perp} {k_\perp^2}& = &{\epsilon_\perp} {\epsilon_\parallel} \ {\omega^2} / {c^2},
\label{eq:dispersion}
\end{eqnarray}
where the subscripts $\parallel$ and $\perp$ represent the directions correspondingly parallel and perpendicular to the symmetry axis. When the real parts of the  two orthogonal components of the permittivity, $\epsilon_\parallel$ and $\epsilon_\perp$, have opposite sign, the corresponding iso-frequency
surface is a hyperboloid (see Fig. 1(a) ), formed by either one (${\rm Re}\left[ \epsilon_\parallel \right] > 0$,  ${\rm Re}\left[ \epsilon_\perp \right]  <  0$) or two sheets (${\rm Re}\left[ \epsilon_\parallel \right] > 0$,  ${\rm Re}\left[ \epsilon_\perp \right]  <  0$) .  As the group velocity (and the Poynting vector) in the low-low limit are orthogonal to the iso-frequency surface, with the hyperbolic surface asymptotically approaching a code for $k \gg \omega/c$, the corresponding electromagnetic wave propagate in the directions at the angle 
\begin{eqnarray}
\vartheta & = & \arctan \sqrt{- \epsilon_\parallel / \epsilon_\perp}
\label{eq:theta}
\end{eqnarray}
with respect to the symmetry axis of the uniaxial material (see Fig. 1(b)). As a result, in a hyperbolic medium light emitter by a ``point" (subwavelength) dipole, forms the narrow conical beam pattern -- see Fig. 1(c).

In the limit of a lossless local hyperbolic medium, the corresponding emission intensity is singular at the emission cone defined by Eqn. (\ref{eq:theta}), however optical absorption, \cite{Vinod} finite size of the material unit cell, \cite{nmat,Vinod,Noginov,Zhang}   and the essential nonlocality of free-carrier electromagnetic response due to their inherent mobility \cite{EN:PRA} remove this singularity.

\section{Long Range Dipole-Dipole Interaction in Hyperbolic Nanostructures}

We begin by considering the setup that, while challenging to realize in real experimental environment,  clearly illustrates the underlying physics and  allows a
straightforward theoretical analysis.
Consider a point emitter at the symmetry axis of a cylindrical waveguide of radius $R$ fabricated from a uniaxial hyperbolic material (such as e.g. sapphire \cite{Sapphire1,Sapphire2}, hBN \cite{hBN1} or a plethora of other natural media\cite{Naturally-hyperbolic} -- see  \ref{fig:HM_dipole}(c). With essentially unlimited \cite{unit-cell} propagating wavenumbers in a hyperbolic medium (see Eqn. (\ref{eq:dispersion}) and Fig. \ref{fig:HM_dipole}(a)), the spatial spectrum of the electromagnetic fields radiated by a point source,  is primarily supported by the waves that are confined to the waveguide by total internal reflection,
which will then lead to the natural ``auto-focusing'' \cite{EN-autofocusing} into multiple focal spots separated by the distance
\begin{eqnarray}
\Delta z & = &   \frac{ 2 m R}{{\rm Re}\left[ \sqrt{-\epsilon_\perp / \epsilon_\parallel} \right]}.
\label{eq:foci}
\end{eqnarray}
The spatial width of these foci $\delta z$ is only limited by atomic scale $a_0$ and the hyperbolic material absorption, as the latter  has a relatively stronger impact on large-wavenumber components:
\begin{eqnarray}
\delta z_m & = & {\rm max}\left\{a_0,   2 m    \frac{{\rm Im}\left[ \sqrt{-\epsilon_\perp / \epsilon_\parallel} \right]}{\left| \epsilon_\perp / \epsilon_\parallel \right|^{3/4}} R \ \right\},
\end{eqnarray}
where $m$ is an integer label of the subsequent focal spots.

The quantum evolution of two atomic dipoles positioned at the axis of a hyperbolic cylindrical resonator and separated by the distance $\Delta z$, is then defined by the effective Hamiltonian
\begin{eqnarray}
\hat{\cal H}_{12} & = & \frac{\hbar \omega_{eg}}{2} \left( \hat{\sigma}_z^{(1)} +\hat{\sigma}_z^{(2)} \right) 
\nonumber \\
& - &
  \frac{J_{12}}{2}  \left(  \hat{\sigma}_x^{(1)} \hat{\sigma}_x^{(2)} + \hat{\sigma}_y^{(1)} \hat{\sigma}_y^{(2)}
 \right), \label{eq:H12} 
\end{eqnarray}
where $\omega_{eg}$ is the quantum transition frequency between the ground ($| g\rangle$) and excited ($| e\rangle$) states of the qubit, with $\omega_{eg}$  in the hyperbolic frequency band of the resonator,
${\bf p} \equiv \langle g | \hat{\bf p} | e \rangle \equiv e {\bf r}_{eg}$ is the dipole moment associated with this transition,
$ \hat{\bf \sigma}_\alpha^{(j)}$ is the Pauli matrix for site $j$, and the interaction energy
 \begin{eqnarray}
J_{12}  =   \frac{p^2}{8 R^3} \frac {{\rm Re} \left[\frac{1}{\epsilon_\perp} \frac{1 + \epsilon_\parallel\epsilon_\perp}{1 - \epsilon_\parallel\epsilon_\perp}\right] }{{\rm Im}  \sqrt{-\epsilon_\parallel/\epsilon_\perp}^3}  \sim  \frac{e^2}{\left| \epsilon\right|r_{eg}} \left( \frac{r_{eg}}{R}\ \frac{\left|\epsilon\right|}{{\rm Im}\left[\epsilon\right]}\right)^3. \ \ \ \ 
\label{eq:J12_1}
 \end{eqnarray}
 With $r_{eg} \sim 1 \ {\rm nm}$ in donor-based quantum emitters, for a macroscopic resonator with the radius $R \simeq 100$ nm fabricated from an isotopically enriched 
hBN \cite{hBNisotope} we find $J_{12} \simeq 100 \ {\rm meV} \simeq \hbar\omega$, which corresponds to the regime of ultra-strong coupling.\cite{strong-coupling-review}

The challenge of actual experimental realization of the system that we just discussed stems from two fundamental issues. First, there is 
 the practical issues of finding a suitable qubit with the high coherence and quantum transitions within the hyperbolic band, that could be ``inserted'' at the axis of a low-loss natural hyperbolic material with nanometer-scale accuracy.  Second problem with this setup, which perhaps is even more important, is that the
 emitter placed in the bulk of a polar crystal,  will inevitably form a scattering center for the optical phonons, thus immediately increasing their scattering rates. This will inevitably result in the proportional rise in the  imaginary part of the dielectric permittivity and the corresponding drastic reduction of the coupling due to its cubic power in Eqns. (\ref{eq:J12_1}).

Therefore, for a practical quantum computing platform based on qubit interactions mediated by  low-loss hyperbolic optical medium, the qubits must be physically separated from the hyperbolic material. Due to the strong impedance mismatch with the hyperbolic medium, such a spacer layer however, will substantially reduce the qubit-to-qubit coupling that is mediated by the hyperbolic material. 

However, strong and long-range  interaction mediated by the optical hyperbolic medium, can be  dramatically enhanced by the phenomenon of the Hyperbolic Super-Resonance. This enhancement can offset the 
reduction of the interaction energy due to the spacer layer separating the qubits from the hyperbolic medium, 
leading to practical Hyperbolic Quantum Processor.  

 \begin{figure*}[htbp] 
 \centering
 \includegraphics[width=7.in]{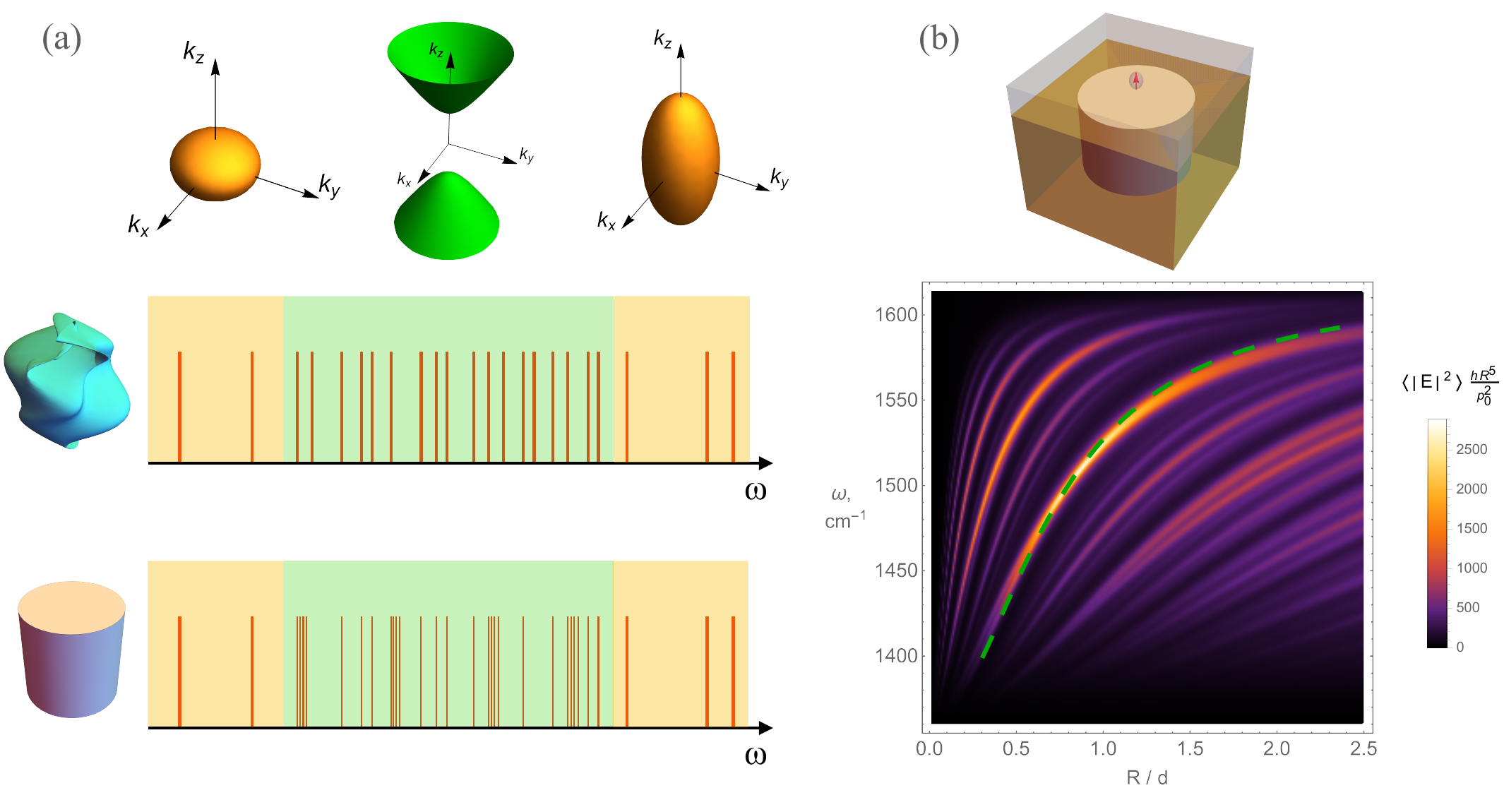} 
   \caption{The  hyperbolic Super-Resonance.  Panel (a): the conceptual introduction.
   Compared to the frequency range with dielectric response (Re$[\epsilon_\parallel \epsilon_\perp]>0$,
   light orange shading), hyperbolic frequency bands   (Re$[\epsilon_\parallel \epsilon_\perp]<0$,
   light green shading) show a dramatically increased mode density. The top row shows the iso-frequency surfaces in the corresponding frequency intervals. 
   In a symmetric resonator  geometry  (top mode density diagram), resonant modes avoid degeneracies via the general avoided crossing mechanism inherent to the underlying non-integrable dynamics.\cite{Gutzwiller,ENProcSPIE,SunFogler2015}. In a symmetric 
   geometry (bottom) such as the cylinder with circular or elliptic cross-section,  regular dynamics leads to level clustering, that results in the  hyperbolic Super-Resonances  with the asymptotically exact {\it infinite} mode degeneracy  in the effective medium limit.  
    Panel (b): the average intensity of the light emitted by a point dipole, that is trapped in the 
   hyperbolic resonator. Top row show  the schematics the system, with a small dipole emitter 
   on top of a cylinder hyperbolic resonator.  Light-orange color corresponds to the hyperbolic material, 
   gold -- to the metal cladding, gray --  to the material supporting the emitter; in a practical physical realization
   of such system, the hyperbolic material is hBN, metal cladding is gold, and the point emitter is 
   a donor impurity atom in silicon. The false color plot shows the average trapped intensity as a function of the resonator aspect ratio and 
   emission frequency. The dielectric permittivity parameters of the hyperbolic medium correspond to hBN. 
   The dashed green line shows the main super-resonance, defined by our Eqn.  (\ref{eq:hSR-cylinder}).
   Note nearly absent light intensity outside the hyperbolic frequency band $1380 \ {\rm cm}^{-1} < \omega < 1620 \ {\rm cm}^{-1}$, and its dramatic increase at the super-resonance.
   }
   \label{fig:hSR}
 \end{figure*}

\section{Hyperbolic Super-Resonance}

One of the key features of a natural hyperbolic material that makes it an ideal
medium to interface with quantum electronic dynamics, is its ability to support
optical frequency waves ultra-high (up to the $X$-ray spatial scale) wavenumbers that can
effectively couple with point-like quantum defects and emitters -- see Fig. \ref{fig:HM_dipole}(a).
This nearly-unlimited range of electromagnetic modes in hyperbolic media manifests in
the broadband ``super-singularity'' of the corresponding photonic density of states, \cite{PDOSprl} 
and leads to strong coupling between hyperbolic polaritons in hBN slabs with
electronic excitations in two dimensional electron systems in close proximity to the hyperbolic medium.
\cite{TomadinPolini2015,RioloPolini2024}

For a finite-size hyperbolic resonator of an arbitrary shape, this super-singularity leads to a dense set of modes in the hyperbolic band,\cite{spacings-comment}
with the total number limited by the natural wavenumber cut-off at
the inverse atomic scale \cite{unit-cell} 
-- see Fig. \ref{fig:hSR}(a). With the right choice of the shape and dimensions of the hyperbolic resonant cavity, one can achieve nearly-perfect (exact in the
limit of large wavenumbers) degeneracy of an  (almost) infinite \cite{unit-cell} number of modes at the same frequency - see Fig. \ref{fig:hSR}(a).

Such infinite mode degeneracies in hyperbolic resonators also arise in other geometries,\cite{EN-hSR} as long as the corresponding ray-optical dynamics remains integrable -- which can be shown in a straightforward manner, using the transformation  to action-angle variables, as in the derivation of Einstein-Brillouin-Keller semiclassical quantization 
procedure.\cite{Gutzwiller} In particular, for the hyperbolic Super-Resonance in a metal-clad cylindrical hyperbolic resonator (see Fig. \ref{fig:hSR}(b)) we obtain:
\begin{eqnarray}
\sqrt{- \frac{\epsilon_\perp\left(\omega_{r}\right)} {\epsilon_\parallel\left(\omega_{r}\right)}} & =  &\frac{4 R}{d} m,
\label{eq:hSR-cylinder}
\end{eqnarray}
where $R$ and $d$ are respectively the radius and the length of the cylindrical resonator, and $m$ is a (positive) integer.  
In Fig.  \ref{fig:hSR}(b) we show the total electromagnetic energy, trapped in the cylindrical hexagonal Boron Nitride resonator, that was radiated by at point dipole in its proximity that is 
oscillating at the frequency in the second hyperbolic band hBN,  in false color as a function of the frequency and the resonator aspect ratio $d/R$. The green line in Fig.  \ref{fig:hSR}(b)  is defined by Eqn. 
(\ref{eq:hSR-cylinder}) for $m = 1$.

In the geometry of \ref{fig:hSR}(b), for a quantum emitter with the transition dipole moment ${\bf p } \equiv p \hat{\bf z}$, at the HSR frequency for the 
corresponding Jaynes-Cummings model \cite{Jaynes-Cummings}  coupling coefficient $g$ we obtain \cite{EN-hSR}
\begin{eqnarray}
\hbar g & = & \frac{1 - \cos\frac{\pi m}{2}}{2 }
\sqrt{\frac{p^2 \hbar \omega}{3 d h^2}},
\label{eq:g}
\end{eqnarray}
where the integer $m$ is the order of the Super-Resonance (see Eqn. (\ref{eq:hSR-cylinder})). For a quantum emitter based on deep donor impurities in silicon, with the characteristic dipole moment $p \sim e\cdot1$ nm, the dipole-resonator spacing $h \simeq 3$ nm, and the cylinder length $d \sim 50$ nm, Eqn. (\ref{eq:g})  
yields $\hbar g \simeq 0.15 \hbar\omega$ for lower-frequency and  $\hbar g \simeq 0.1 \hbar\omega$  for the higher-frequency hyperbolic bands in hexagonal boron nitride. Thus, at the Hyperbolic Super-Resonance, a single quantum emitter in the proximity to a hyperbolic optical resonator, naturally operates in the regime of strong coupling.
 
\begin{figure}[htbp] 
 \centering
  \includegraphics[width=3.in]{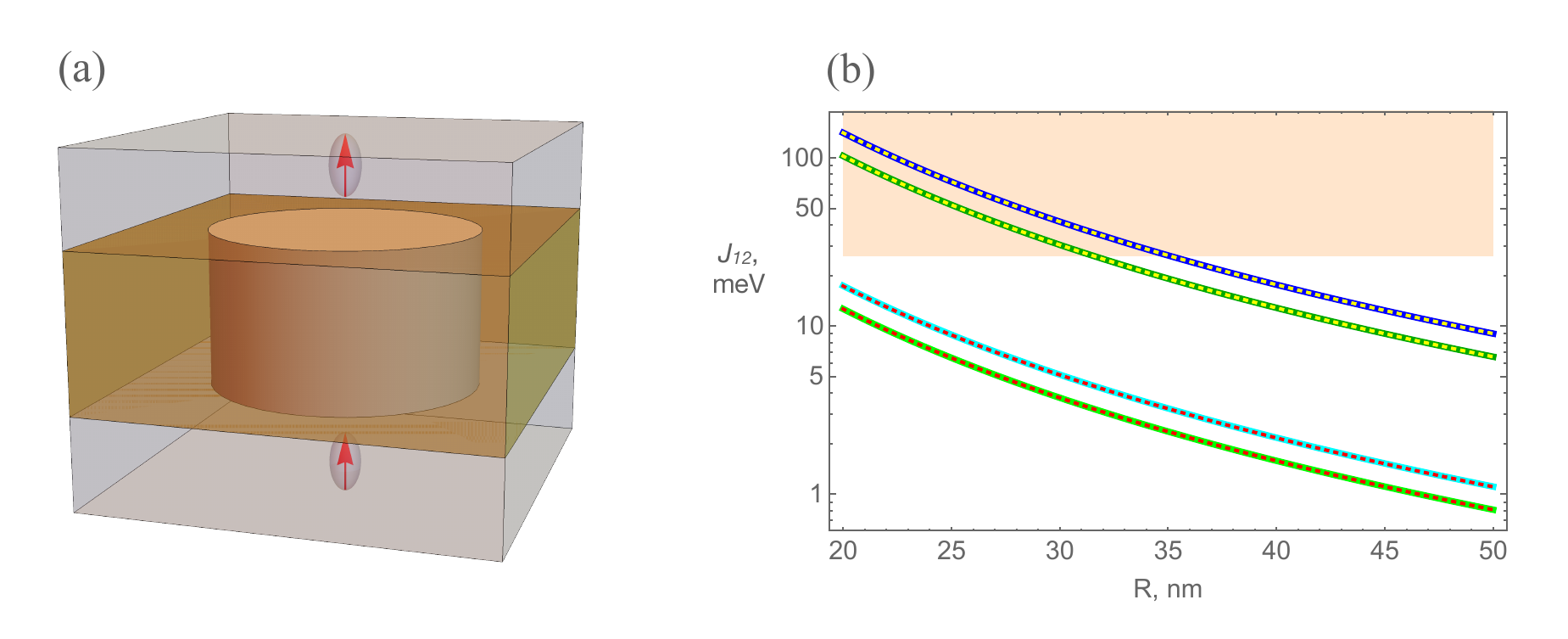} 
   \caption{Long-range dipole-dipole interaction mediated by hyperbolic medium, at the Super-Resonance. Panel (a): the schematics of 
   a macroscopic  hyperbolic cylindrical resonator (with the same color coding as in Fig. \ref{fig:hSR}), now with two 
   oscillating dipole emitters in its near field.  Panel (b): the dipole-dipole interaction energy $J_{12}$ as a function of the radius of the resonator,
   at the conditions of the first- (blue curves) and second-order (green lines) hyperbolic Super-Resonance.   The top and bottom 
   pairs of the curves correspond to the higher- and lower-frequency hyperbolic bands in { natural}  hBN (respectively $\hbar\omega \sim 100$ meV and $\hbar\omega \sim 180$ meV). The shaded are corresponds to the energy ranbge above the room tem[perature thermal energy ($\simeq 22$ meV). Note that
   in the isotopically enriched hBN the interaction energy $J_{12}$ is higher by a factor of $ \sim 3$ \cite{hBN4} and is therefore
   comparable to the emitted photon energy, so that the system is  in the ultra-strong coupling limit $J \simeq \hbar\omega$. }
   \label{fig:hSR_2dipoles}
 \end{figure}

 For two  oscillating dipoles ${\bf p}_1$ and ${\bf p}_2$ on the opposites sides of a cylindrical hyperbolic resonator in
 the geometry of  Fig. \ref{fig:hSR_2dipoles}, for the interaction energy at the hyperbolic Super-Resonance we obtain
at the frequency of the hyperbolic Super-Resonance, we  obtain
\begin{eqnarray}
J_{12} & \simeq & \frac{8  p^2 }{ h_*^3 + 2 h^3 } ,
\label{eq:J12}
\end{eqnarray}
where
\begin{eqnarray}
h_* \equiv d \left|  {\rm Im}\sqrt{- \frac{\epsilon_\perp}{\epsilon_\parallel}} \right|.
\end{eqnarray}
For the ``extended'' hyperbolic resonator, formed by 
``intersecting'' elliptic resonators ( see  Fig. \ref{fig:HQP}(b)), the interaction energy is
given by
\begin{eqnarray}
J'_{ij} & = & \frac{1 - e_h^2}{2 e_h} J_{ij},
\end{eqnarray} 
where $e_h$ is the eccentricity of the ellipse of  the cross-section geometry of the hyperbolic resonator,
and $J_{ij}$ is defined by Eqn. (\ref{eq:J12}).

In Fig. \ref{fig:hSR_2dipoles}(b), for two quantum emitters with the quantum transition dipole matrix element $p = e\cdot 1 \ {\rm nm}$ 
(which is the scale relevant to $1s \to 2 p$  transitions in donor atoms in silicon \cite{Ramdas}), separated 
by the distance of $h = 5 \ {\rm nm}$ from the  hexagonal boron nitride  resonator interface, we plot the corresponding interaction energy as a function of the cylinder radius. The atomic transition frequency $\omega$ corresponds to the Hyperbolic Super-Resonance in the higher-frequency hyperbolic band of $h$BN, 
and  the pink shaded area shows the range where the quantum emitter interaction energy exceeds the 
room temperature thermal energy $kT \simeq 0.1 \hbar \omega_0$. Note that the quantum emitters are strongly coupled ($J > 0.1 \hbar \omega$ \cite{strong-coupling-review})
for all but the largest separation distance ($R > 50$ nm). It is this inherently strong quantum dipole-dipole coupling that is the foundation for the proposed Hyperbolic Quantum Processor.

\section{Quantum Evolution and Qubit Entanglement Mediated by Hyperbolic Media}

\begin{figure}[htbp] 
 \centering
 \includegraphics[width=3.5in]{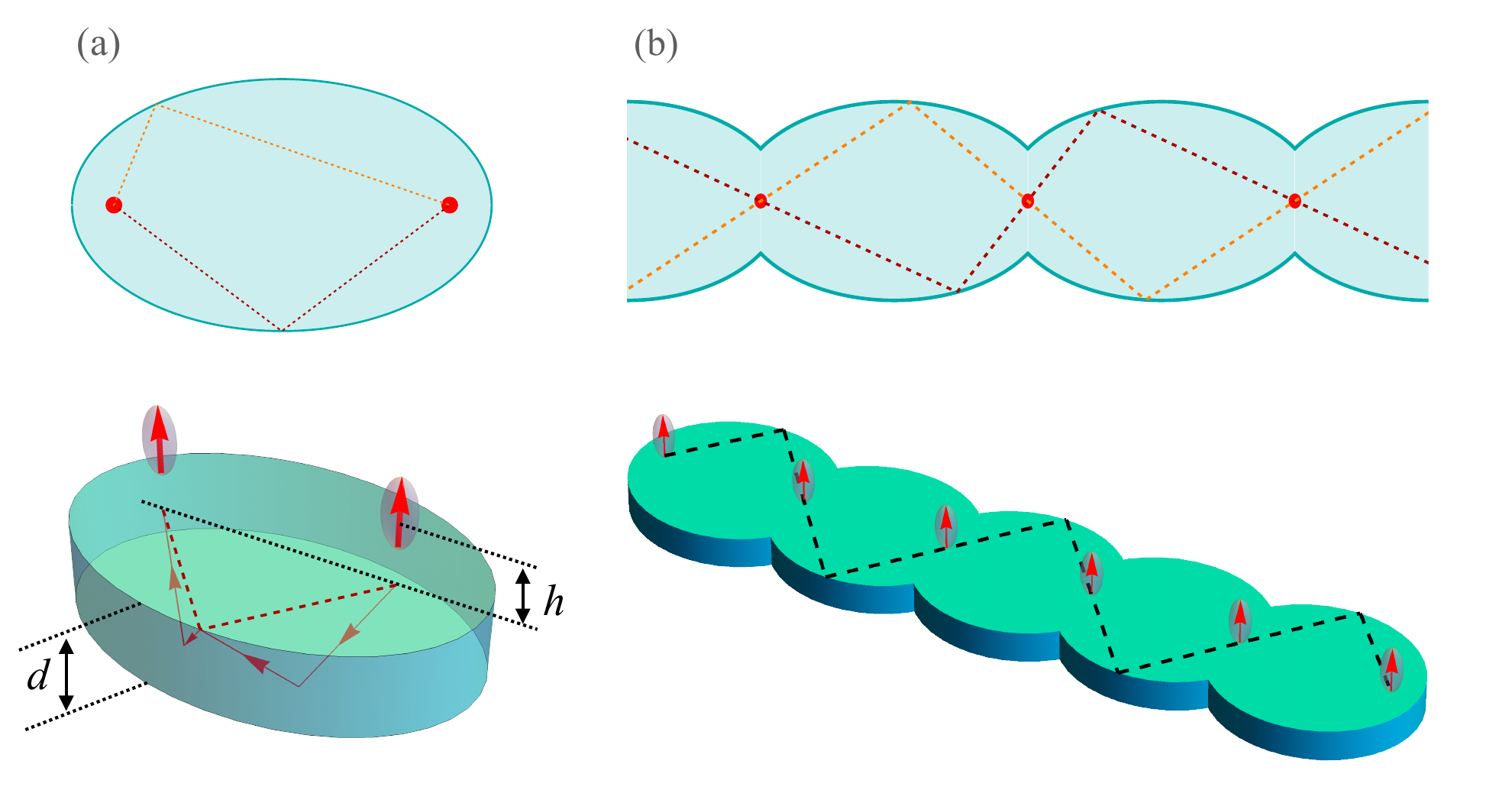} 
   \caption{Panel (a): two quantum emitters  (e.g. chalcogene double donor in silicon) are positioned above the foci of an hyperbolic (e.g. hBN) resonator in the elliptical cylinder geometry. With the light in the hyperbolic medium not subject to diffraction broadening, photons from one emitter are focused to the other, leading to strong coupling at macroscopic separation.Panel (b): the architecture of the Hyperbolic Quantum processor, with  atomic qubits at the foci of the multi-elliptic resonator with hyperbolic dispersion. The quantum emitters are represented by red circles (top) and red arrows (bottom). Dashed lines show the projections of the 3D ray trajectories (solid red lines with arrows) in the hyperbolic resonator onto its top surface. Black dotted lines in (a) are used to indicate the height of the resonator $d$ and the spacing thickness $b$ between the resonator and the qubits.
}
   \label{fig:HQP}
 \end{figure}

To describe the quantum interaction between the donor atom cubits in the silicon layer coupled to the hyperbolic resonator (see Fig. \ref{fig:HQP}), we will
use the Dyadic Green function \cite{DyadicGF}- based  theoretical formalism of Refs. \cite{Asenjo-GarciaKimble2017,Asenjo-GarciaChang2017} based 
on the early work of Welsch and colleagues. \cite{GrunerWelsch1996,DungWelsch2002,BuhmannWelsch2007} While the original approach of \cite{Asenjo-GarciaKimble2017,Asenjo-GarciaChang2017}  was applied to the optical response of atoms and other quantum emitters coupled to one-dimensional photonic structures such as photonic crystal waveguides, it is in fact general and only relies on the broad spectrum of the Green’s function -- which is perfectly suitable to a hyperbolic medium with its singularly high photonic density of states in an optically broad bandwidth. Then, the donor atoms qubit density matrix $\hat{\rho}_A$ evolution is defined by the 
generalized  von Neumann-Liouvillle  equation
\begin{eqnarray}
\frac{\partial\hat{\rho}_A}{\partial t} & = & \frac{i}{\hbar} \left[\hat{\rho}_A, \hat{\cal H}\right]  +\hat{\cal L}\left[\hat{\rho}_A\right],
\end{eqnarray}
where, within the rotating wave approximation, and in the frame rotating with the probe field frequency, the Hamiltonian and Lindblad operators are
\begin{eqnarray}
\hat{\cal H} & = & \sum_j \frac{\hbar \omega_j}{2} \hat{\sigma}^j - 
\sum_{i,j} \theta_i \theta_j J_{ij} \hat{\sigma}_{eg}^i \hat{\sigma}_{ge}^j \label{eq:H} \\
& + & \sum_j \left( {\bf p}_j^*\cdot{\bf E}\left({\bf r}_j\right) e^{- i \Delta_j t} \hat{\sigma}^j_{eg} +  {\bf p}_j\cdot{\bf E}^*\left({\bf r}_j\right) e^{i \Delta_j t} \hat{\sigma}^j_{eg}  \right), 
 \nonumber  \\
\hat{\cal L} & = & \sum_{i,j} \left( \theta_i \theta_j \Gamma_{ij} + \delta_{ij} \left(1 - \theta_j\right) \gamma_j \right) \nonumber \\
& \times & \left( 2 \sigma_{ge}^i \hat{\rho}_A\sigma_{eg}^j - \sigma_{eg}^i \sigma_{ge}^j \hat{\rho}_A  - \hat{\rho}_A   \sigma_{eg}^i \sigma_{ge}^j \right).
\label{eq:Lindblad}
\end{eqnarray}
Here the variables $\theta_j \in \{0, 1\}$ define whether the qubit transition frequency (tuned by dynamic Stark  shifts due to 
off-resonance optical control fields -- see Fig. \ref{fig:HQPaction}(b)~)  is in the hyperbolic  ($\theta = 1$)  or in the dielectric ($\theta = 0$) frequency band, and $\gamma_j \ll \Gamma_{jj}$ is the decoherence rate in the dielectric regime. Note that due to the macroscopic separation between different qubits, the effective spin-exchange coupling  is only present between different quantum emitters when their transition frequencies are both shifted to the hyperbolic frequency band. 

The electric field ${\bf E}$ in the effective Hamiltonian (\ref{eq:H}) corresponds to the on-resonance optical control
at the frequency $\omega$ close to the qubit transition frequency $\omega_j$ between its ground ($| g\rangle$) and excited ($| e\rangle$) states, 
${\bf p} \equiv \langle g | \hat{\bf p} | e \rangle$ is the dipole moment associated with this transition,
$\Delta_j = \omega_j - \omega$ is the detuning between the optical field and the transition in qubit $j$,  
$\hat{\sigma}_{eg}^j = | e \rangle \langle g |$
is the atomic coherence operator  between the ground and excited states of atom $j$ ,  and 
$\hat{\sigma}^j =   | e \rangle \langle e | -  | g \rangle \langle g |$. In terms of the Pauli matrices, $\hat{\sigma} \equiv \sigma_z$,  $\hat{\sigma}_{ge} \equiv (\sigma_x - i \sigma_y)/2$,  $\hat{\sigma}_{eg} \equiv (\sigma_x + i \sigma_y)/2$. The effective spin-exchange and decay rates \cite{Asenjo-GarciaKimble2017,Asenjo-GarciaChang2017} are
\begin{eqnarray}
J_{ij} + i \ \Gamma_{ij} & = & 4 \pi \left(\frac{\omega}{c}\right)^2 {\bf p}_i^*\cdot{\bf G}\left({\bf r}_i, {\bf r}_j, \omega\right)\cdot {\bf p}_j,
\label{eq:JG}
\end{eqnarray}
where ${\bf G}$ is the dyadic Green's function of the system without the dipole emitters.\cite{DyadicGF,Asenjo-GarciaKimble2017}

Due to the hyperbolic Super-Resonance, the effective spin-exchange coupling $J^{ij}$ is comparable to the optical photon energy ($\hbar \omega \sim 100$ meV, corresponding to the optical period $T = 2 \pi/\omega \sim 0.04 $ ps) -- see Fig. \ref{fig:hSR_2dipoles}, and exceeds the atomic decoherence rates by several orders of magnitude. In particular, at the $4$K  temperature the excited donor states in isotopically pure silicon show the coherence times in excess of 300 picoseconds, which are three orders of magnitude larger then the corresponding transition frequencies.\cite{GreenlandAeppli2010}
Even at the liquid nitrogen temperatures, experiments in doped silicon have shown \cite{VinhPNAS2008}  the lifetimes approaching $200$ ps.

However, the hyperbolic Super-Resonance also leads to the increased decoherence due to material absorption in the hyperbolic resonator, described by the  decay rates 
$\Gamma^{ij}$ in Eqn. (\ref{eq:Lindblad}), so that to turn the system of donor atom cubits coupled to a hyperbolic resonator into a viable platform for quantum computation, 
we need 
\begin{eqnarray}
J^{ij} \gg \Gamma
\end{eqnarray}
for all different $\Gamma_{ij}$.
 At the hyperbolic Super-Resonance condition we find
 \begin{eqnarray}
 \Gamma_{i \neq j} \ll \Gamma_{ii},
 \end{eqnarray}
 where
 \begin{eqnarray}
\Gamma_{ii}  
& \simeq & 
\frac{p^2}{h^3} \ 
\frac{3 h_*}{\sqrt{h^2 + 9 h_*^2}} 
 \left|  {\rm Re}\left[\left(\frac{\epsilon_\perp}{\epsilon_\parallel}\right)^2 \left(1 - \frac{\epsilon_\perp}{\epsilon_\parallel}\right)
\right] \right|,
\label{eq:Gamma11}
 \end{eqnarray}
 
 Note that, even in the simple geometry of Fig. \ref{fig:hSR_2dipoles}, the exchange coupling energy $J_{12}$ and  the decoherence rates  $\Gamma_{11} \equiv \Gamma_{22}$,  show very contrasting  behavior with the variation of the spacer thickness $h$. This qualitative difference is especially pronounced in the small
 $h$ limit, when $\Gamma_{11}$ diverges and $J_{12}$ remains finite (see Eqns. (\ref{eq:J12}) and (\ref{eq:Gamma11})). With a finite amount of loss, while propagating over the distance separating different emitter sites, the ultra-high wavenumber field components  are absorbed, leaving only a finite amount in the resulting exchange coupling energy. On the other hand, in the limit of $h\to 0$, the emitter is coupled to all wavenumber components supported  by the hyperbolic medium, leading to the well-known divergence of the decoherence rate for a point emitter in a continuous hyperbolic  medium.\cite{PDOSapl,ScienceTT}
 
As a result of this essential difference in the variation of $J_{12}\left(h\right)$ and $\Gamma_{11}\left(h\right)$,    for 
 $h \ll h_*$ in the geometry of Fig. \ref{fig:hSR_2dipoles},  we find that $J_{12} < \Gamma_{11}$, while in the opposite 
 limit, $h\gtrsim h_*$, from Eqns. (\ref{eq:J12}) and (\ref{eq:Gamma11})
\begin{eqnarray}
\frac{\Gamma_{11}}{J_{12}} & \sim &
  \frac{h_*}{h},
\label{eq:J}
\end{eqnarray}
so to ensure that $J \gg \Gamma$ we need the spacer thickness $h$ between the qubit and the hyperbolic resonator, to {\it exceed} the critical value $h_* \equiv d \left|  {\rm Im}\sqrt{- {\epsilon_\perp}/{\epsilon_\parallel}} \right|$:
\begin{eqnarray} 
h \gg h_*  .
\label{eq:loss}
\end{eqnarray}
However, to ensure fast and efficient gate  operations that involve multi-qubit entanglement, we need the effective exchange energy close to the strong-coupling  limit,
\begin{eqnarray}
J^{ij} \gtrsim 0.1 \ \hbar\omega,
\end{eqnarray}
which implies 
\begin{eqnarray}
h \lesssim h_c \equiv 40 \sqrt[3]{\frac{e^2 r_{eg}^2}{\hbar\omega}}.
\label{eq:coupling}
\end{eqnarray}
From  (\ref{eq:loss}) and (\ref{eq:coupling}) we therefore obtain the strong inequality
\begin{eqnarray}
h_* \ll h \lesssim h_c. 
\label{eq:thickness_range}
\end{eqnarray}
While the spacer thickness $h$ can be set to the desired value at the fabrication level, the strong inequality (\ref{eq:thickness_range}) 
can only be satisfied when
\begin{eqnarray}
\frac{h_c}{h_*} \equiv \frac{40}{\left| {\rm Im}  \sqrt{\epsilon_\perp/{\epsilon_\parallel}}\right|} \ \frac{r_{eg}}{d} 
\  \sqrt[3]{\frac{e^2 / r_{eg}}{\hbar\omega}} \gg 1.
\end{eqnarray}

The essential feature of the electromagnetic response in high quality polar crystals such as e.g. sapphire \cite{Sapphire1,Sapphire2} or hexagonal boron nitride \cite{hBN1} 
is the relatively small loss in their hyperbolic bands. In particular, for natural hBN at room temperature the experimental values of the measured dielectric permittivity
tensor components yield ${\rm Im}  \sqrt{\epsilon_\perp/{\epsilon_\parallel}} \simeq 0.03$. Furthermore, in isotopically enriched hBN samples due to the elimination of  phonon scattering on isotopically different impurities the absorption is substantially smaller,  with experimental values of ${\rm Im}  \sqrt{\epsilon_\perp/{\epsilon_\parallel}} \simeq 0.01$. Moreover, as the dominant phonon scattering mechanism in isotopically pure high quality crystals is 
phonon-phonon scattering, the resulting scattering rates are proportional to temperature, so that in the cryogenic regimes the optical absorption  in the hyperbolic bands of hBN and other polar crystals will be even lower. As a result, at liquid nitrogen temperatures and below, essential to preserve the electronic coherence of donor atoms in silicon chosen as out qubit platform, for $d \sim 50$ nm and $r_{eg} \sim 2$ nm, we find
\begin{eqnarray}
\frac{h_c}{h_*} \simeq 100,
\end{eqnarray}
which with the proper choise of the spacer thickness $h$ from (\ref{eq:thickness_range}) is more than sufficient to simultaneously satisfy the requirements of low decoherence and strong optical coupling.

Note that, with the choice of $h \simeq  h_c $ resulting in $\Gamma \lesssim J/100$, for the corresponding gate operation we find the fidelity \cite{fidelity} on the order of $99\%$ and above. The gate fidelity in the proposed Hyperbolic Quantum Processor based on donor impurity qubits in a silicon layer deposited onto the hexagonal boron nitride resonator in the ``multi-elliptical'' geometry (see Figs. \ref{fig:HQPaction} and \ref{fig:HQP}(b)) can be even higher, possibly reaching $99.9\%$ and beyond, with the use of isotopically pure crystalline hBN layer with atomically-flat interfaces at lower temperatures, leading to the reduced
absorption of hyperbolic phonon-polaritons dominated by phonon-on-phonon scattering.\cite{Ziman}

With the addition of separate  plasmonic waveguides that separately reach the near fields of each of the donor atom qubits and allow for the individual optical control,  in e.g. the technologically well developed nano-coaxial geometry,\cite{nano-waveguide} the ``multi-elliptical''
system of Fig.  \ref{fig:HQPaction} and Fig. \ref{fig:HQP}(b) forms the proposed Hyperbolic Quantum Processor.

\begin{figure*}[htbp] 
 \centering
 \includegraphics[width=7.in]{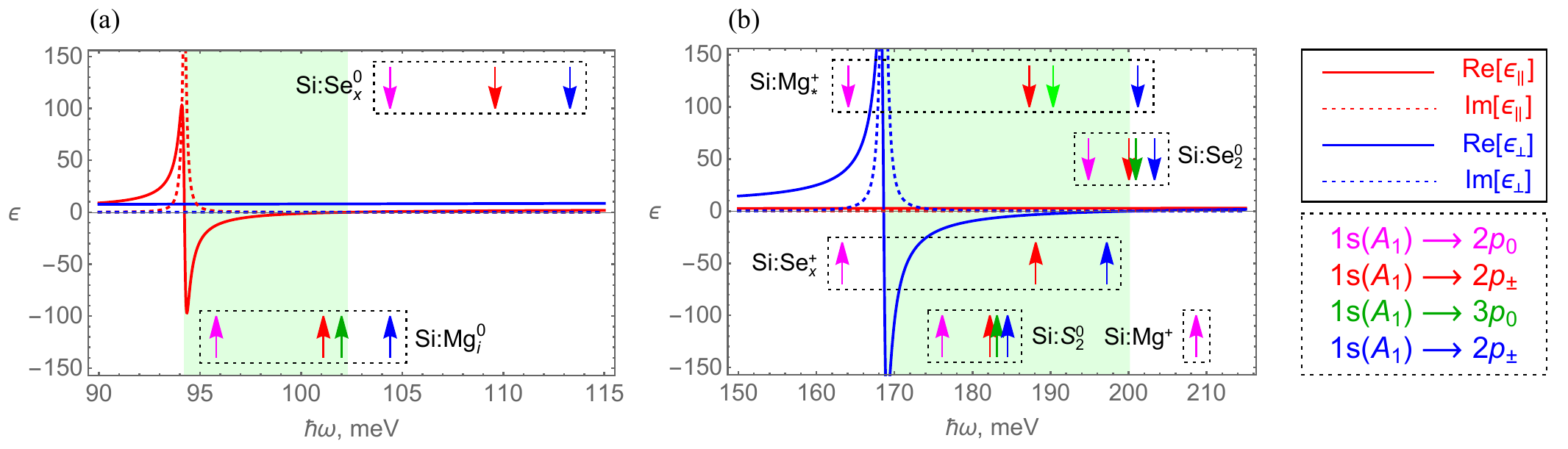} 
   \caption{The dielectric permittivity of the natural hexagonal boron nitride in its lower-frequency (a) and higher-frequency (b) hyperbolic bands. 
   Blue and red lines correspond respectively to $\epsilon_\parallel\left(\omega\right)$ and  $\epsilon_\perp\left(\omega\right)$, with the real and
   imaginary parts represented by solid and dotted curves. Light-green shading indicates the frequency bands with the hyperbolic dispersion.
   Arrows show the photon emission frequencies for different donor atoms in silicon, where the color coding (defined in the legend on the right)
   represents specific atomic transitions. Note that multiple well-known donor impurities show the electronic transitions in or close to the 
   hyperbolic bands of the hexagonal boron nitride.  
    }
   \label{fig:hBN_emitters}
 \end{figure*}

\section{Quantum Emitters for Low-Loss Hyperbolic Materials}

It has now been recognized that donor impurities in silicon offer the solid state analogue of long-lived Rydberg states in atoms and ions. 
Similar to the electrons of the outer shells in isolated atoms at an ion trap, the loosely bound donor electrons orbit the charged atomic cores
in the effective ``vacuum" of the conducting band -- leading to the electron spectra characteristic of isolated atoms and ions. 
Manipulation of the Rydberg-like donors impurities in semiconductors by single and multiphoton processes has demonstrated,\cite{VinhPNAS2008} and
the dominant decoherence mechanism for their excited states in isotopically purified silicon was shown to  lifetime broadening,\cite{VinhPNAS2008} just 
as for atoms in ion traps.\cite{QC3} Furthermore, the donors can be effectively isolated from the environment and have no significant sources of decoherence other than population decay by emission of phonons.\cite{VinhPNAS2008}

A electron in a donor atom in silicon, naturally bound by its Coulomb potential, therefore offers an excellent cubit, \cite{StempNatComm2024} 
 with demonstrated single-qubit gate fidelity up to 99.98\%.\cite{DehollainNJPhys2016} However, electron two-qubit logic gates based on donor atoms in silicon face the major
 challenge that their coupling rapidly decays with the inter-donor distance, leading to the effective interaction range on the order of  $10$ nm \cite{KoillerPRL2001,VoisinNatComm2020,JoeckerNJPhys2021} and
 making it nearly impossible to control multi-cubit entanglement while being able to separately access the individual qubits.\cite{StempNatComm2024} 
 
As we have demonstrated in this work,  qubits embedded in or positioned at the direct proximity of a hyperbolic material waveguide, show 
long range dipole-dipole interactions and, due to  phenomenon of the Hyperbolic Super-Resonance,
can reach strong coupling regime at optical frequencies even at macroscopic qubit separation.  
As a result,
qubits formed by donor atoms embedded in the silicon layer at the surface of a hyperbolic optical resonator or waveguide, can be  both individually 
accessed and entangled together, either at the  two- and multi-qubit level,  by external electromagnetic fields at mid-infrared frequencies. As follows from Eqn. (\ref{eq:H}), the quantum evolution in such Hyperbolic Quantum Processor (HQP)  corresponds to the standard Ising spin model with
optically controlled on-site energies and long-range exchange interaction.

Moreover, all the necessary requirements for the actual physical implementation of the scalable HQP
only rely on the capabilities to grow low-loss hyperbolic materials (such as e.g. hexagonal Boron Nitride \cite{hBN1,hBN2,hBN3,hBN4}), deposit silicon layers on the surface of these materials, \cite{hBN3,hBN4} choose between multiple different  donor atoms that support long-lifetime Rydberg-like states with transitions at the frequencies within or in close proximity to the hyperbolic bands in low-loss hyperbolic materials, \cite{SwartzThomas1980,PajotNaud1985,KlevermanGrimmeiss1986,GreenlandAeppli2010,Ludwig1965,JanzenGrimmeiss1984,StegerAger2009,AbrahamThewalt2018,KaraiskajCardona2003,SwartzThomas19800}   
accurately position donor atoms in silicon with nanometer-scale accuracy, \cite{donor-placement1,donor-placement2} and
fabricate low-loss plasmonic waveguides to individually couple to separate donor qubits -- that have already been experimentally demonstrated. 

In particular, for the effective couple to the hyperbolic bands in high quality polar materials such as hBN (see Fig. \ref{fig:hBN_emitters} for the relevant frequency intervals), we propose to exploit the electric dipole–allowed optical transitions available to ``deep'' donors, such as the chalcogen double donors sulfur, magnesium and selenium. In their neutral state, these helium-like double donors bind two electrons, with larger binding energies in comparison to shallow donors such as phosphorous.. When singly ionized, the remaining electron has an even larger binding energy and a hydrogen-like orbital structure with optical transitions in the mid-infrared (mid-IR). As shown in Fig. \ref{fig:hBN_emitters}, many of these transitions lie within or close to the hyperbolic bands of the hBN. 

Furthermore, a number of these have the desired property of having the frequency just above the hBN hyperbolic bands. In particular, for the lower-frequency hyperbolic band in hBN, we can use $1s(A_1) \to 2p_0$ transition  of the selenium ${\rm Se}_x^0$ donor complex, or the $1s(A_1) \to 2p_\pm$ transition in the magnesium donor atom. On the other hand, for high-frequency hyperbolic band in hBN the desired properties are shown by the single-ionized magnesium donors ${\rm Mg}^+$ ($1s(A_1) \to 2p_0$  transition) and   ${\rm Mg}_*^+$ ($1s(A_1) \to 2p_\pm$  transition). 

\section{Fundamental and Technological Limits}

The proposed Hyperbolic Quantum Processor presents a major fabrication challenge, especially in the highly 
desired limit of large integrated quantum circuit entangling multiple qubits on a single chip. However, every
single technology, necessary for its actual construction, has already been independently demonstrated.  The hexagonal 
boron nitride is now routinely integrated into quantum nanophotonic devices,\cite{SakibShcherbakov2024} and
the atomic level  quality of its surfaces was essential for use together with graphene within a single device.\cite{FukamachiAgo2023} hBN  can be grown to form high quality interface with silicon,\cite{DaiBasov2015} and
many chalcogen atoms - based  deep donor impurities in silicon are known with the 
optical transitions in immediate proximity to both of the hyperbolic bands of hBN -- see Fig.  \ref{fig:hBN_emitters}. Furthermore, in the two decades - long quest for spin qubits in silicon quantum dots, a remarkable set of experimental tools was developed allowing controlled placement of donor atoms in silicon with sub-nanometer accuracy.\cite{BroomeSimmons2018} While integrating these so far separate processes and techniques within a single fabrication pipeline is in no way straightforward, it should be possible within the limits of the existing technology.

The performance of any quantum processor is ultimately limited by the decoherence rates in the system. 
For the hyperbolic phonon polaritons that mediate the qubit interactions in multi-qubit gate operations,  the improvements in the fabrication  quality of hBN by e.g. isotope purification that removes the phonon decoherence  due to isotope scattering and the reduction of the system temperature that ``freezes out" the multi-phonon scattering processes,\cite{Ziman} can naturally bring the resulting multi-qubit gate operations fidelity to $99.$\% and beyond. It is therefore the 
decoherence rate of the electronic states of the donor qubits in silicon layers that will define limits to the performance of the proposed Hyperbolic Quantum Processor.

What fundamentally separates the proposed here approach from the existing alternatives, from superconducting \cite{Ezratty2023} to  trapped ions \cite{ionQ} to neutral atoms,\cite{EveredLukin2023}  is that, being essentially a 
silicon - based optoelectronic chip, the Hyperbolic Quantum Processor offers
a direct pathway to the high-density qubit integration, thus taking the full advantage of the quantum error correction \cite{Shor1995} to bring the elusive dream of practical quantum computing to the realm of practical optoelectronics.

Furthermore, a Hyperbolic Quantum Processor can be used as a simulation platform for quantum magnetic systems,\cite{HungKimble2016} while its inherently strong light-matter coupling offers a direct pathway to the realization of 
 optical nonlinearities at the level of individual photons in the IR frequency range.\cite{ChangLukin2007,FirstenbergVuletic2013,ChangLukin2014,ChangKimble2018,ZhangPohl2022}

\section{Acknowledgements}

E.D. would like to acknowledge discussions with D. Basov, A. Imamoglu, J. Feng and S. Chattopadhyay, and funding from SNSF (project 200021\_212899), Swiss State Secretariat for Education, Research and Innovation (contract number UeM019-1) and NCCR SPIN (a National Centre of Competence in Research, funded by  Swiss National Science Foundation, grant number 225153)

$$ \ $$ 

\section*{Supplementary Material}
\beginsupplement

\subsection*{Electrodynamics in Hyperbolic Media}

From the late 19th century when optical interferometers brought the dawn of modern physics, offering a peak into the nature and the chemical composition of distant galaxies by spectroscopic analysis of 
starlight and leading to the birth of Theory of Relativity through the Michelson-Morley experiment,  to the present time 
when optical sensors are ubiquitous in all aspects of modern world and optoelectronic devices enable most of the communications and data transmissons on Earth, the fundamental physics of the interference of propagating electromagnetic waves has been the cornerstone of the applied optics and engineering. While extremely accurate and 
robust, interference-based conventional optical devices in their spatial dimensions must exceed, or at least be comparable to  the optical wavelength defined by the speed of light and the 
optical frequency in order to build up strong constructive and destructive interference -- which puts a fundamental roadblock on the route to miniaturization and integration of optoelectronic systems.

It was the realization that hyperbolic materials, supporting propagating waves with essentially unlimited wavenumbers,\cite{PN,hyperlens1,hyperlens2} that recently opened the way to genuine nanophotonic devices, whose spatial dimensions can be orders of magnitude below the free-space optical wavelength. From the fundamental physics point of view, this strongly subwavelength range that's inherent to all meaningful applications of hyperbolic materials, corresponds to the 
quasistatic  regime \cite{Cai} when Maxwell's equations reduce to
\begin{eqnarray}
\nabla \left[ {\bf \epsilon}\left(\omega, {\bf r}\right)  \nabla \varphi\left(\omega, {\bf r}\right) \right] & = & 4 \pi \rho\left(\omega, {\bf r}\right),  \label{eq:Poisson} \\
{\bf E}\left(\omega, {\bf r}\right) & = & - \nabla\varphi\left(\omega, {\bf r}\right), \label{eq:E-phi}
\end{eqnarray} 
where $\rho\left(\omega, {\bf r}\right)$ is the charge density and $\phi\left(\omega, {\bf r}\right)$ is the scalar potential.
For a point dipole in an infinite  uniaxial medium, from (\ref{eq:Poisson}),(\ref{eq:E-phi})  we immediately obtain
\begin{eqnarray}
{\bf E} & = & - \frac{1}{\epsilon_\parallel \sqrt{\frac{\epsilon_\parallel}{\epsilon_\perp}}}{\bf \nabla}\left({\bf p \cdot\nabla}  
\frac{1}{\sqrt{\rho^2 + \frac{\epsilon_\perp}{\epsilon_\parallel} z^2}}
\right), \label{eq:e_dipole}
\end{eqnarray}
where ${\bf p}$ is the dipole moment of the (point) emitter, while $z$ and $\rho$  correspond to the distance from the emitter in the directions respectively parallel and orthogonal to the symmetry axis of the material. Note that for a hyperbolic medium (when ${\rm Re}\left[ \epsilon_\parallel \epsilon_\perp\right] < 0$) Eqn. (\ref{eq:e_dipole}) shows the characteristic conical emission pattern of Fig. \ref{fig:HM_dipole} and the divergence of the field in the lossless limit when $z \to \rho \tan\vartheta \equiv \rho \ \sqrt{{ - \epsilon_\parallel}/{\epsilon_\perp}}$. 

The quantum evolution dynamics of the Hyperbolic Quantum Processor, set by the effective Hamiltonian (\ref{eq:H}),
is controlled by the effective spin exchange energies 
$J_{ij}$ and related decay rates  $\Gamma_{ij}$, defined by the quantum dipole moments of individual qubits and the 
dyadic Green function ${\bf G}\left({\bf r}_i, {\bf r}_j, \omega\right)$ of the classical electromagnetic theory -- see Eqn. (\ref{eq:JG}). For nanostructures in nontrivial geometry, formed by anisotropic media,  in the general case ${\bf G}$ does not have a simple analytical representation, which does not allow a direct calculation of the spin-exchange coupling and associated decay rates via Eqn. (\ref{eq:JG}).

However, Eqn. (\ref{eq:JG}) can be equivalently expressed as
\begin{eqnarray}
J_{ij} + i \Gamma_{ij} & = & {\bf p}_i^* \cdot {\bf E}_j\left({\bf r}_i,\omega\right),
\label{eq:JG0SM}
\end{eqnarray}
where ${\bf E}_j\left({\bf r},\omega\right)$ is the electric field of the oscillating point dipole ${\bf p}_i$, located at 
${\bf r}_i$.  Because of the inherently quasistatic regime of classical electrodynamics in hyperbolic nanostructures, 
which dramatically reduces the mathematical complexity of the problem of the full set of Maxwell's equations in nontrivial geometry to the solution of  Eqns. (\ref{eq:Poisson}),(\ref{eq:E-phi}), the calculation of the electric field of a 
single point dipole ${\bf E}_j\left({\bf r}_i, \omega\right)$ for Eqn. (\ref{eq:JG0SM}) now becomes straightforward, leading to explicit analytical expressions for $J_{ij}$ and 
$\Gamma_{ij}$.

\subsection*{Spin-Exchange Coupling and Decoherence}

With the qubits defined by deep donor impurities in a semiconductor such as silicon, the inherent 
anisotropy of the crystalline environment breaks the energy level degeneracy originally associated with the spherically symmetric atom, so that the dipole-allowed transitions when the dipole moment parallel (${\bf p} \parallel \hat{\bf z}$) or when it is perpendicular  (${\bf p} \perp \hat{\bf z}$ ) to the symmetry axis, now have very different frequencies -- see 
Fig. \ref{fig:hBN_emitters}. We therefore only need to consider the spin-exchange coupling when dipoles are either both parallel or both perpendicular the $z$-axis.

In the geometry of Fig. \ref{fig:hSR_2dipoles}, for the two  oscillating dipoles ${\bf p}_1$ and ${\bf p}_2$ on the opposite sides of a cylindrical hyperbolic resonator,  we obtain
\begin{eqnarray}
 {\bf p}_1^* \cdot {\bf E}_2\left({\bf r}_1,\omega\right) & = & -  
\frac{2 \pi p_1^* p_2}{R^2} 
\sum_{n=1}^\infty
\frac{x_{on}^2 \ \exp\left( - x_{0n} \frac{h}{R}\right)}{D_n}, \ \ \ \
\label{eq:V_2dipoles}
\end{eqnarray}
where
\begin{eqnarray}
D_n & = & \cos\left(\sqrt{-\frac{\epsilon_\perp}{\epsilon_\parallel} }x_{0n} \frac{d}{E}\right)
+ \frac{1}{2} \left( \frac{\epsilon_0}{\epsilon_\perp} \sqrt{\-\frac{\epsilon_\perp}{\epsilon_\parallel}} \right.
\nonumber \\
& - & \left.  \frac{\epsilon_\perp}{\epsilon_0} \sqrt{-\frac{\epsilon_\parallel}{\epsilon_\perp}} \right)
\sin\left(\sqrt{-\frac{\epsilon_\perp}{\epsilon_\parallel} }x_{0n} \frac{d}{E}\right),
\end{eqnarray}
$x_{0n} \simeq \pi\left( n - 1/4\right)$ is the $n$-th zero of the Bessel function $J_0\left(x\right)$, and $\epsilon_0$ is the dielectric permittivity of the spacer layers suppporting the dipoles, with the pole corresponding to the Hyperbolic Super-Resonance frequency.
Note that Eqn. (\ref{eq:V_2dipoles}) can be also expressed as
\begin{eqnarray}
{\bf p}_1^* & \cdot & {\bf E}_2\left({\bf r}_1,\omega\right)   =  4 \pi  
\frac{p_1p_2}{R^2} 
\frac{1}{1 + \frac{i}{2} \left( \frac{\epsilon_0}{\epsilon_\perp} \sqrt{\-\frac{\epsilon_\perp}{\epsilon_\parallel}} - \frac{\epsilon_\perp}{\epsilon_0} \sqrt{-\frac{\epsilon_\parallel}{\epsilon_\perp}} \right)}
 \nonumber \\
& \times & 
\sum_{n=1}^\infty
\frac{x_{0n}^2 \ \exp\left( x_{0n} \left( i \sqrt{-\frac{\epsilon_\perp}{\epsilon_\parallel}} \frac{d}{R} - \frac{h}{R}\right)\right)}{1 - r_{\rm eff} 
\exp\left( 2 i \sqrt{-\frac{\epsilon_\perp}{\epsilon_\parallel}}  x_{0n}\frac{d}{R} \right)
}, \ \ \ \ \ \ \ \ \ \ 
\label{eq:V_2dipoles_r}
\end{eqnarray}
where the effective reflective coefficient
\begin{eqnarray}
r_{\rm eff} & = & \left(-1\right)^{2 n} \ \left( - \frac{1 + i \frac{\epsilon_0}{\epsilon_\parallel} \sqrt{-\frac{\epsilon_\parallel}{\epsilon_\perp}}}{1 - i \frac{\epsilon_0}{\epsilon_\parallel} \sqrt{-\frac{\epsilon_\parallel}{\epsilon_\perp}} }\right)^2
\label{eq:reff}
\end{eqnarray}
is the product of the reflection coefficients at the ``bounce points"  of the periodic ray trajectory
in the hyperbolic resonator connecting the dipoles. 
 
 At the frequency of the primary ($m=1$) hyperbolic Super-Resonance defined by Eqn. (\ref{eq:hSR-cylinder}), we then obtain
\begin{eqnarray}
J_{12} & \simeq & \frac{4 \ p^2}{h^3 + 32 \left(  \left| \frac{{\rm Im} \sqrt{{-\epsilon_\perp}/{\epsilon_\parallel}}}{{\rm Re}\sqrt{{-\epsilon_\perp}/{\epsilon_\parallel}}}\right|  R\right)^3}, 
\label{eq:SMJ12}
\end{eqnarray}
 and
 \begin{eqnarray}
\Gamma_{11} & = &  \Gamma_{22}  = \frac{p^2}{2 h^3} \ 
 {\rm Re}\left[\frac{\epsilon_0}{\epsilon_\parallel^2}
\sqrt{-\frac{\epsilon_\perp}{\epsilon_\parallel}}
\left(\epsilon_\perp - \epsilon_\parallel\right)
\right]  \nonumber \\
& \times & \int_0^\infty dt \ t^2 e^{-t}  \tanh\left({\rm Im}\sqrt{- \frac{\epsilon_\perp}{\epsilon_\parallel}}
\frac{d}{h}\ t
\right) \nonumber \\
& \simeq & 
\frac{p^2}{h^3} \ 
\frac{\left|  {\rm Re}\left[\frac{\epsilon_0 \left(\epsilon_\perp - \epsilon_\parallel\right)}{\epsilon_\parallel^2}
\sqrt{-\frac{\epsilon_\perp}{\epsilon_\parallel}}
\right] \right|}{\sqrt{1 + \frac{h^2}{\left( 3  \ d  \ {\rm Im}\sqrt{- {\epsilon_\perp}/{\epsilon_\parallel}}  \right)^2 } }}.
\label{eq:SMGamma11}
 \end{eqnarray}

\end{document}